\documentclass[12pt]{revtex4} %Updated Sep 24, 2009%
\usepackage{amssymb,epsf}
\usepackage{latexsym}
\begin{document}
\title{\bf  Dark pressure in a non-compact and non-Ricci flat 5D Kaluza-Klein cosmology}
\author{F. Darabi\thanks{e-mail:
f.darabi@azaruniv.edu}\\{\small Department of Physics, Azarbaijan
University of Tarbiat Moallem, 53714-161, Tabriz, Iran } }
\vspace{200mm}

\begin{abstract}
\textbf{{\bf \center{Abstract:}}} \\
In the framework of noncompact Kaluza-Klein theory, we
investigate a $(4+1)$-dimensional universe consisting of a
$(4+1)$ dimensional Robertson-Walker type metric coupled to a
$(4+1)$ dimensional energy-momentum tensor. The matter part
consists of an energy density together with a pressure subject to
$4D$ part of the $(4+1)$ dimensional energy-momentum tensor. The
dark part consists of just a dark pressure $\bar{p}$,
corresponding to the extra-dimension endowed by a scalar field,
with no element of dark energy. It is shown that for a flat
universe, coupled with the non-vacuum states of the scalar field,
the reduced field equations subject to suitable equations of
state for matter and dark part may reveal inflationary behavior
at early universe, deceleration for radiation dominant era and
then acceleration in matter dominant era.
\end{abstract}
\maketitle
\newpage
\section{INTRODUCTION}

The recent distance measurements from the light-curves of several
hundred type Ia supernovae \cite{1,2} and independently from
observations of the cosmic microwave background (CMB) by the WMAP
satellite \cite{3} and other CMB experiments \cite{4,5} suggest
strongly that our universe is currently undergoing a period of
acceleration. This accelerating expansion is generally believed to
be driven by an energy source called dark energy. The question of
dark energy and the accelerating universe has been therefore the
focus of a large amount of activities in recent years. Dark energy
and the accelerating universe have been discussed extensively from
various point of views over the past few years
\cite{Quintessence,Phantom,K-essence}. In principle, a natural
candidate for dark energy could be a small positive cosmological
constant. This is usually studied in $\Lambda$CDM model. This model
attempts to explain: cosmic microwave background observations, large
scale structure observations, and supernovae observations of
accelerating universe. It uses Friedmann-Robertson-Walker (FRW)
metric, the Friedmann equations and the cosmological equations of
state to describe the universe from right after the inflationary
epoch to the present and future times, according to Einstein
equation with a cosmological constant
$$
R_{\mu \nu}-\frac{1}{2}Rg_{\mu \nu}=kT_{\mu \nu}+\Lambda g_{\mu
\nu}.
$$
Alternative approaches have also been pursued, a few example of
which can be found in \cite{sahni,cardassian,chaplygin}. These
schemes aim to improve the quintessence approach overcoming the
problem of scalar field potential, generating a dynamical source
for dark energy as an intrinsic feature. Quintessence is a scalar
field with an  equation of state which unlike cosmological
constant, varies through the space and time. Many models of
quintessence have a tracker behavior. In these models, the
quintessence field has a density which closely tracks (but is
less than) the radiation density until matter-radiation equality
, which triggers the quintessence to start, having
characteristics similar to dark energy and eventually dominating
the universe and starting the acceleration of the universe. The
goal would be to obtain a comprehensive model capable of linking
the picture of the early universe to the one observed today, that
is, a model derived from some effective theory of quantum gravity
which, through an inflationary period would result in today
accelerated Friedmann expansion driven by some
$\Omega_{\Lambda}$-term.

Another approach in this direction is to employ what is known as
modified gravity where an arbitrary function of the Ricci scalar,
namely $f(R)$, is inserted into the Einstein-Hilbert action
instead of Ricci scalar $R$ as
$$
S=-\frac{1}{2k}\int d^4x \sqrt{-g}f(R) +S_m(g_{\mu \nu}, \psi),
$$
where $\psi$ is the matter field. This  is  a  fully covariant
theory based on the principle of least action. One of the main
considerations with this modified gravity is to know how can we
fit this theory with the local solar system tests as well as
cosmological constraints \cite{solar,chameleon}. It has been
shown that such a modification may account for the late time
acceleration and the initial inflationary period in the evolution
of the universe so that positive powers of $R$ lead to inflation
while negative powers of $R$ result in current acceleration
\cite{modifidgravity2,modifidgravity,Deffayet}. A scenario where
the issue of cosmic acceleration in the framework of higher order
theories of gravity in $4D$ is addressed can be found in
\cite{capo}. One of the first proposals in this regard was
suggested in second reference of \cite{modifidgravity} where a
term of the form $R^{-1}$ was added to the usual Einstein-Hilbert
action.

The idea that our world may have more than four dimensions is due
to Kaluza \cite{Kaluza}, who unified Einstein's theory of General
Relativity with Maxwell's theory of Electromagnetism in a $5D$
manifold. In 1926, Klein reconsidered Kaluza's idea and treated
the extra dimension as a topologically compact small circle
\cite{Klein}. Since then the Kaluza-Klein idea has been studied
extensively from different angles. Notable amongst them is the
so-called Space-Time-Matter (STM) theory, proposed by Wesson and
his collaborators, which is designed to explain the origin of
matter in terms of the geometry of the bulk space in which our
$4D$ world is embedded, for reviews see \cite{Wesson}. More
precisely, in STM theory, our world is a hypersurface embedded in
a five-dimensional Ricci-flat ($R_{AB}=0$) manifold where all the
matter in our world can be thought of as being the manifestation
of the geometrical properties of the higher dimensional space
according to
$$
G_{\alpha \beta}=8 \pi  T_{\alpha \beta}.
$$
Applications of the idea of induced matter or induced geometry can
also be found in other situations \cite{FSS}. The STM theory allows
for the metric components to be dependent on the extra dimension and
does not require the extra dimension to be compact. The sort of
cosmologies stemming from STM theory is studied in
\cite{LiuW,STM-cosmology,WLX,LiuM,Ponce}. The evolution of the
universe has also been studied extensively based on this noncompact
{\it vacuum} Kaluza-Klein theory \cite{Bellini} where a 5D mechanism
is developed to explain, by a single scalar field, the evolution of
the universe including inflationary expansion and the present day
observed accelerated expansion.

Since in a variety of inflationary models the scalar fields have
been used in describing the transition from the quasi-exponential
expansion of the early universe to a power law expansion, it is
natural to try to understand the present acceleration of the
universe by constructing models where the matter responsible for
such behavior is also represented by a scalar field. Such models
are worked out, for example, in Ref \cite{6}. In this chapter,
based on the above mentioned ideas on higher dimension and scalar
field, a 5$D$ cosmological model is introduced which is not Ricci
flat, but is extended to be coupled to a higher dimensional
energy momentum tensor. It is shown that the higher dimensional
sector of this model may induce a dark pressure, through a scalar
field, in four dimensional universe so that for a flat universe
under specific conditions one may have early inflation and current
acceleration even for the non-vacuum states of the scalar field.

\section{Space-Time-Matter versus Kaluza-Klein theory}

The Kaluza-Klein theory is essentially general relativity in 5$D$
subject to two conditions: \\
1) the so called ``cylinder'' condition, introduced by Kaluza, to
set all partial derivations with respect to the 5$^{th}$ coordinate
to zero\\
2) the ``compactification'' condition, introduced by Klein, to set a
small size and a closed topology for the 5$^{th}$ coordinate.\\
Physically, both conditions have the motivation of explaining why we
perceive the 4 dimensions of space-time and do not observe the fifth
dimension. In perfect analogy with general relativity, one may
define a $5\times 5$ metric tensor $g_{AB}\:(A,B=0, 1, 2, 3, 4)$
where 4 denotes the extra coordinate which is referred to
``internal'' coordinate. Moreover, one can form a 5$D$ Ricci tensor
$R_{AB}$, a 5$D$ Ricci scalar $R$ and a 5$D$ Einstein tensor
$G_{AB}=R_{AB}-\frac{1}{2}Rg_{AB}$. In Principle, the field
equations are expected to be
\begin{equation}
G_{AB}=kT_{AB},\label{00}
\end{equation}
where $k$ is an appropriate coupling constant and $T_{AB}$ is a 5$D$
momentum tensor. In Kaluza-Klein theory, however, it is usually
assumed that the universe in 5$D$ is empty, so we have the {\it
vacuum} field equations which may equivalently be defined as
\begin{equation}
R_{AB}=0,\label{111}
\end{equation}
where $R_{AB}$ is the 5$D$ Ricci tensor. These 15 relations serve to
determine the 15 components of the metric $g_{AB}$. To this end,
some assumptions are to be made on $g_{AB}=g_{AB}(x^C)$ as the
choice of coordinates or gauges. Interested in electromagnetism and
its unification with gravity, Kaluza realized that $g_{AB}$ may be
expressed in the following form that involves the electromagnetic
4-potentials $A_{\alpha}$ as well as an scalar field $\Phi$
\begin{equation}
g_{AB}=\left(\begin{array}{cc} (g_{\alpha \beta}-\kappa^2 \Phi^2 A_{\alpha}A_{\beta}) & -\kappa \Phi^2 A_{\alpha}\\
\\-\kappa \Phi^2 A_{\beta} &  -\Phi^2 \end{array}\right)_.\label{22-}
\end{equation}
The five dimensional Ricci tensor and Christoffel symbols exactly as
in four dimension are defined in terms of the metric as
\begin{equation}
R_{AB}= \partial_C \Gamma^C_{AB}-\partial_B \Gamma^C_{AC}+
\Gamma^C_{AB}\Gamma^D_{CD}-\Gamma^C_{AD}\Gamma^D_{BC},\label{220}
\end{equation}
\begin{equation}
\Gamma^C_{AB}=\frac{1}{2}g^{CD}(\partial_A g_{DB}+
\partial_B g_{DA}-\partial_D g_{AB}).\label{221}
\end{equation}
Then the field equations $R_{AB}=0$ reduces to
\begin{equation}
G_{\alpha \beta}=\frac{\kappa^2 \Phi^2}{2}T_{\alpha
\beta}-\frac{1}{\Phi}(\nabla_{\alpha}\nabla_{\beta}\Phi-g_{\alpha
\beta}\Box\Phi),\label{330}
\end{equation}
\begin{equation}
\nabla^{\alpha}F_{\alpha
\beta}=-3\frac{\nabla^{\alpha}\Phi}{\Phi}F_{\alpha \beta},\label{44}
\end{equation}
\begin{equation}
\Box\Phi=-\frac{\kappa^2\Phi^3}{4}F_{\alpha \beta}F^{\alpha
\beta},\label{55}
\end{equation}
where $G_{\alpha \beta}$, $F_{\alpha \beta}$ and $T_{\alpha \beta}$
are the usual 4$D$ Einstein tensor, electromagnetic field strength
tensor and energy-momentum tensor for the electromagnetic field,
respectively. The Kaluza's case $\Phi^2=1$ together with the
identification $\kappa=(16 \pi G)^{\frac{1}{2}}$ leads to the
Einstein-Maxwell equations
\begin{equation}
G_{\alpha \beta}={8\pi G}T_{\alpha \beta},\label{66}
\end{equation}
\begin{equation}
\nabla^{\alpha}F_{\alpha \beta}=0.\label{77}
\end{equation}
The STM or ``induced matter theory'' is also based on the vacuum
field equation in 5$D$ as $R_{AB}=0$ or $G_{AB}=0$. But, it differs
mainly from the Kaluza-Klein theory about the cylinder condition so
that one may now keep all terms containing partial derivatives with
respect to the fifth coordinate. This results in the 4$D$ Einstein
equations
\begin{equation}
G_{\alpha \beta}=8\pi G T_{\alpha \beta},\label{88}
\end{equation}
provided an appropriate definition is made for the energy-momentum
tensor of matter in terms of the extra part of the geometry.
Physically, the picture behind this interpretation is that curvature
in $(4+1)$ space induces effective properties of matter in $(3+1)$
space-time. The fact that such an embedding can be done is supported
by Campbell's theorem \cite{Compbell} which states that any
analytical solution of the Einstein field equations in $N$
dimensions can be locally embedded in a Ricci-flat manifold in
$\left(N+1 \right)$ dimensions. Since the matter is induced from the
extra dimension, this theory is also called the {\it induced matter
theory}. The main point in the induced matter theory is that these
equations are a subset of $G_{AB}=0$ with an effective or induced
4$D$ energy-momentum tensor $T_{\alpha \beta}$, constructed by the
geometry of higher dimension, which contains the classical
properties of matter.

Taking the metric (\ref{22-}) and choosing coordinates such that the
four components of the gauge fields $A_{\alpha}$ vanish, then the
5-dimensional metric becomes
\begin{equation}
g_{AB}=\left(\begin{array}{cc} g_{\alpha \beta} & 0\\
\\0 &  \epsilon \Phi^2 \end{array}\right)_,\label{222}
\end{equation}
where the factor $\epsilon$ with the requirement $\epsilon^2=1$ is
introduced in order to allow a timelike, as well as spacelike
signature for the fifth dimension. Using the definitions
(\ref{220}), (\ref{221}) and keeping derivatives with respect to the
fifth dimension, the resultant expressions for the $\alpha\beta$,
$\alpha4$ and $44$ components of the five dimensional Ricci tensor
$R_{AB}$ are obtained
$$
\hat{R}_{\alpha \beta}={R}_{\alpha
\beta}-\frac{\nabla_{\beta}(\partial_{\alpha}\Phi)}{\Phi}
\hspace{200mm}
$$
\begin{equation}
\hspace{20mm}+ \frac{\epsilon}{2\Phi^2}\left(\frac{\partial_{4}\Phi
\partial_{4}g_{\alpha \beta}}{\Phi}-\partial_{4}g_{\alpha \beta}+
g^{\gamma \delta}\partial_{4}g_{\alpha \gamma}\partial_{4}g_{\beta \delta}
-\frac{g^{\gamma \delta}\partial_{4}g_{\gamma
\delta}\partial_{4}g_{\alpha \beta}}{2}\right),
\end{equation}
$$
\hat{R}_{\alpha 4}=\frac{g^{44}g^{\beta
\gamma}}{4}+(\partial_{4}g_{\beta \gamma}
\partial_{\alpha}g_{44}-\partial_{\gamma}g_{44}
\partial_{4}g_{\alpha \beta})\hspace{200mm}
$$
$$
+\frac{\partial_{\beta}g^{\beta \gamma}
\partial_{4}g_{\gamma \alpha}}{2}+\frac{g^{\beta \gamma}
\partial_{4}(\partial_{\beta}g_{\gamma \alpha})}{2}-\frac{\partial_{\alpha}g^{\beta \gamma}
\partial_{4}g_{\beta \gamma}}{2}
$$
\begin{equation}
\hspace{50mm}-\frac{g^{\beta \gamma}
\partial_{4}(\partial_{\alpha}g_{\beta \gamma})}{2}+
\frac{g^{\beta \gamma}g^{\delta \epsilon}\partial_{4}g_{\gamma
\alpha}\partial_{\beta}g_{\delta \epsilon}}{4} +
\frac{\partial_{4}g^{\beta \gamma}\partial_{\alpha}g_{\beta
\gamma}}{4},
\end{equation}
$$
\hat{R}_{44}=-\epsilon \Phi\Box \Phi- \frac{\partial_{4}g^{\alpha
\beta}\partial_{4}g_{\alpha \beta}}{2}-\frac{g^{\alpha \beta}
\partial_{4}(\partial_{4}g_{\alpha \beta})}{2}\hspace{200mm}
$$
\begin{equation}
\hspace{50mm}+\frac{\partial_{4}\Phi g^{\alpha
\beta}\partial_{4}g_{\alpha \beta}}{2\Phi}- \frac{g^{\alpha
\beta}g^{\gamma \delta }\partial_{4}g_{\gamma
\beta}\partial_{4}g_{\alpha \delta}}{4}.
\end{equation}
Assuming that there is no higher dimensional matter, the Einstein
equations take the form $R_{AB}=0$ which produces the following
expressions for the 4-dimensional Ricci tensor
\begin{equation}
{R}_{\alpha
\beta}=\frac{\nabla_{\beta}(\partial_{\alpha}\Phi)}{\Phi}-
\frac{\epsilon}{2\Phi^2}\left(\frac{\partial_{4}\Phi
\partial_{4}g_{\alpha \beta}}{\Phi}-\partial_{4}g_{\alpha \beta}+
g^{\gamma \delta}\partial_{4}g_{\alpha \gamma}\partial_{4}g_{\beta
\delta} -\frac{g^{\gamma \delta}\partial_{4}g_{\gamma
\delta}\partial_{4}g_{\alpha \beta}}{2}\right),
\end{equation}
\begin{equation}
\nabla_{\beta}\left[\frac{1}{2\sqrt{\hat{g}_{44}}}(g^{\beta
\gamma}\partial_4 g_{\gamma \delta}-\delta^{\beta}_{\alpha}g^{\gamma
\epsilon}\partial_4 g_{\gamma \epsilon})\right]=0,
\end{equation}
\begin{equation}
\epsilon \Phi\Box \Phi=- \frac{\partial_{4}g^{\alpha
\beta}\partial_{4}g_{\alpha \beta}}{4}-\frac{g^{\alpha \beta}
\partial_{4}(\partial_{4}g_{\alpha \beta})}{2}+\frac{\partial_{4}\Phi g^{\alpha
\beta}\partial_{4}g_{\alpha \beta}}{2\Phi}.
\end{equation}
The first equation introduces an induced energy-momentum tensor as
\begin{equation}
8\pi G T_{\alpha \beta}=R_{\alpha \beta}-\frac{1}{2}Rg_{\alpha
\beta}.\label{R}
\end{equation}
The four dimensional Ricci scalar is obtained
\begin{equation}
R=g^{\alpha \beta}R_{\alpha
\beta}=\frac{\epsilon}{4\Phi^2}[\partial_{4}g^{\alpha
\beta}\partial_{4}g_{\alpha \beta}+(g^{\alpha
\beta}\partial_{4}g_{\alpha \beta})^2].
\end{equation}
Inserting $R$ and $R_{\alpha \beta}$ into eq.(\ref{R}) one finds
that
$$
8\pi G T_{\alpha
\beta}=\frac{\nabla_{\beta}(\partial_{\alpha}\Phi)}{\Phi}\hspace{200mm}
$$
\begin{equation}
-\frac{\epsilon}{2\Phi^2}\left[\frac{\partial_{4}\Phi
\partial_{4}g_{\alpha \beta}}{\Phi}-\partial_{4}g_{\alpha \beta}+
g^{\gamma \delta}\partial_{4}g_{\alpha \gamma}\partial_{4}g_{\beta
\delta} -\frac{g^{\gamma \delta}\partial_{4}g_{\gamma
\delta}\partial_{4}g_{\alpha \beta}}{2}+\frac{g_{\alpha
\beta}}{4}(\partial_{4}g^{\gamma \delta}\partial_{4}g_{\gamma
\delta}+(g^{\gamma \delta}\partial_{4}g_{\gamma \delta})^2)\right].
\end{equation}
Therefore, the 4-dimensional Einstein equations $G_{\alpha
\beta}=8\pi G T_{\alpha \beta}$ are automatically contained in the
5-dimensional vacuum equations $G_{AB}=0$, so that the matter
$T_{\alpha \beta}$ is a manifestation of pure geometry in the higher
dimensional world and satisfies the appropriate requirements: it is
symmetric and reduces to the expected limit when the cylinder
condition is re-applied.

\section{THE EXTENDED MODEL}

As explained above, both Kaluza-Klein and STM theories use the
vacuum field equations in 5 dimensions. In Kaluza-Klein case, the
energy-momentum tensor is limited to Electromagnetic and scalar
fields. In STM case, $T_{\alpha \beta}$ includes more types of
matter but is limited to those obtained for an specific form of the
5$D$ metric. We are now interested in a 5$D$ model with a general
5$D$ energy-momentum tensor which is, in principle, independent of
the geometry of higher dimension and can be set by physical
considerations on 5-dimensional matter. We start with the $5D$ line
element
\begin{equation}
dS^2=g_{AB}dx^Adx^B. \label{0}
\end{equation}
The space-time part of the metric $g_{\alpha \beta}=g_{\alpha
\beta}(x^{\alpha})$ is assumed to define the Robertson-Walker line
element
\begin{equation}
ds^2=dt^2-a^2(t)\left(\frac{dr^2}{(1-Kr^2)}+r^2(d\theta^2+\sin^2\theta
d\phi^2 )\right), \label{1}
\end{equation}
where $K$ takes the values $+1, 0, -1$ according to a close, flat or
open universe, respectively. We also take the followings
$$
g_{4 \alpha}=0, \:\:\:\: g_{4 4}=\epsilon\Phi^2(x^{\alpha}),
$$
where $\epsilon^2=1$ and the signature of the higher dimensional
part of the metric is left general. This choice has been made
because any fully covariant $5D$ theory has five coordinate degrees
of freedom which can lead to considerable algebraic simplification,
without loss of generality. Unlike the noncompact vacuum
Kaluza-Klein theory, we will assume the fully covariant $5D$
non-vacuum Einstein equation
\begin{equation}
G_{AB}= 8 \pi G T_{AB}, \label{2}
\end{equation}
where $G_{AB}$ and $T_{AB}$ are the $5D$ Einstein tensor and
energy-momentum tensor, respectively. Note that the $5D$
gravitational constant has been fixed to be the same value as the
$4D$ one. In the
following we use the geometric reduction from 5$D$ to 4$D$ as
appeared in \cite{Ponce}. The $5D$ Ricci tensor is given in terms of
the $5D$ Christoffel symbols by
\begin{equation}
R_{AB}= \partial_C \Gamma^C_{AB}-\partial_B \Gamma^C_{AC}+
\Gamma^C_{AB}\Gamma^D_{CD}-\Gamma^C_{AD}\Gamma^D_{BC}. \label{3}
\end{equation}
The $4D$ part of the $5D$ quantity is obtained by putting $A
\rightarrow \alpha$, $B \rightarrow \beta$ in (\ref{3}) and
expanding the summed terms on the r.h.s by letting $C \rightarrow
\lambda, 4$ etc. Therefore, we have
\begin{equation}
\hat{R}_{\alpha \beta}=\partial_{\lambda} \Gamma^{\lambda}_{\alpha
\beta}+\partial_4 \Gamma^4_{\alpha \beta}-\partial_{\beta}
\Gamma^{\lambda}_{\alpha \lambda}-\partial_{\beta} \Gamma^4_{\alpha
4}+ \Gamma^{\lambda}_{\alpha \beta}\Gamma^{\mu}_{\lambda
\mu}+\Gamma^{\lambda}_{\alpha \beta}\Gamma^{4}_{\lambda
4}+\Gamma^{4}_{\alpha \beta}\Gamma^{D}_{4 D}-\Gamma^{\mu}_{\alpha
\lambda}\Gamma^{\lambda}_{\beta \mu}-\Gamma^{4}_{\alpha
\lambda}\Gamma^{\lambda}_{\beta 4}-\Gamma^{D}_{\alpha
4}\Gamma^{4}_{\beta D},\label{4}
\end{equation}
where $\hat{}$ denotes the $4D$ part of the $5D$ quantities. One
finds the $4D$ Ricci tensor as a part of this equation which may be
cast in the following form
\begin{equation}
\hat{R}_{\alpha \beta}={R}_{\alpha \beta}+\partial_4
\Gamma^4_{\alpha \beta}-\partial_{\beta} \Gamma^4_{\alpha 4}
+\Gamma^{\lambda}_{\alpha \beta}\Gamma^{4}_{\lambda
4}+\Gamma^{4}_{\alpha \beta}\Gamma^{D}_{4 D}-\Gamma^{4}_{\alpha
\lambda}\Gamma^{\lambda}_{\beta 4}-\Gamma^{D}_{\alpha
4}\Gamma^{4}_{\beta D}.\label{5}
\end{equation}
Evaluating the Christoffel symbols for the metric $g_{AB}$ gives
\begin{equation}
\hat{R}_{\alpha \beta}={R}_{\alpha
\beta}-\frac{\nabla_{\alpha}\nabla_{\beta}\Phi}{\Phi}.\label{6}
\end{equation}
Putting $A=4, B=4$ and expanding with $C \rightarrow \lambda, 4$ in
Eq.(\ref{3}) we obtain
\begin{equation}
{R}_{4 4}=\partial_{\lambda} \Gamma^{\lambda}_{4 4}-\partial_{4}
\Gamma^{\lambda}_{4 \lambda}+ \Gamma^{\lambda}_{4
4}\Gamma^{\mu}_{\lambda \mu}+\Gamma^{4}_{4 4}\Gamma^{\mu}_{4
\mu}-\Gamma^{\lambda}_{4 \mu}\Gamma^{\mu}_{4 \lambda}-\Gamma^{4}_{4
\mu}\Gamma^{\mu}_{4 4}.\label{7}
\end{equation}
Evaluating the corresponding Christoffel symbols in Eq.(\ref{7})
leads to
\begin{equation}
{R}_{4 4}=-\epsilon\Phi \Box \Phi. \label{8}
\end{equation}
We now construct the space-time components of the Einstein tensor
$$
G_{AB}=R_{AB}-\frac{1}{2}g_{AB}R_{(5)}.
$$
In so doing, we first obtain the $5D$ Ricci scalar $R_{(5)}$ as
$$
R_{(5)}=g^{AB}R_{AB}= \hat{g}^{\alpha \beta} \hat{R}_{\alpha\beta}+
g^{44}R_{44}= g^{\alpha \beta}(R_{\alpha
\beta}-\frac{\nabla_{\alpha}\nabla_{\beta}\Phi}{\Phi})+\frac{\epsilon}{\Phi^2}(-\epsilon\Phi
\Box \Phi)
$$
\begin{equation}
=R-\frac{2}{\Phi}\Box\Phi,\label{9}
\end{equation}
where the $\alpha 4$ terms vanish and $R$ is the $4D$ Ricci scalar.
The space-time components of the Einstein tensor is written
$\hat{G}_{\alpha \beta}=\hat{R}_{\alpha
\beta}-\frac{1}{2}\hat{g}_{\alpha \beta}R_{(5)}$. Substituting
$\hat{R}_{\alpha \beta}$ and $R_{(5)}$ into the space-time
components of the Einstein tensor gives
\begin{equation}
\hat{G}_{\alpha \beta}={G}_{\alpha \beta}+\frac{1}{\Phi}(g_{\alpha
\beta} \Box \Phi- \nabla_{\alpha}\nabla_{\beta}\Phi). \label{10}
\end{equation}
In the same way, the 4-4 component is written ${G}_{4 4 }={R}_{4 4
}-\frac{1}{2}g_{4 4}R_{(5)}$, and substituting ${R}_{4 4}$,
$R_{(5)}$ into this component of the Einstein tensor gives
\begin{equation}
G_{4 4}=-\frac{1}{2}\epsilon R\Phi^2. \label{11}
\end{equation}
We now consider the $5D$ energy-momentum tensor. The form of
energy-momentum tensor is dictated by Einstein's equations and by
the symmetries of the metric (\ref{1}). Therefore, we may assume a
perfect fluid with nonvanishing elements
\begin{equation}
{T}_{\alpha \beta}=(\rho+p){u}_{\alpha} {u}_{\beta}-p{g}_{\alpha
\beta}, \label{12}
\end{equation}
\begin{equation}
{T}_{44}=-\bar{p}g_{44}= -\epsilon\bar{p}\Phi^2, \label{14}
\end{equation}
where $\rho$ and $p$ are the conventional density and pressure of
perfect fluid in the $4D$ standard cosmology and $\bar{p}$ acts as a
pressure living along the higher dimensional sector. Hence, the
field equations (\ref{2}) are to be viewed as {\it constraints} on
the simultaneous geometric and physical choices of $G_{AB}$ and
$T_{AB}$ components, respectively.

Substituting the energy-momentum components (\ref{12}), (\ref{14})
in front of the $4D$ and extra dimensional part of Einstein tensors
(\ref{10}) and (\ref{11}), respectively, we obtain the field
equations
\begin{equation}
G_{\alpha \beta}=8 \pi G [(\rho+p)u_{\alpha} u_{\beta}-pg_{\alpha
\beta}]+\frac{1}{\Phi}\left[\nabla_{\alpha}\nabla_{\beta}\Phi-\Box
\Phi g_{\alpha \beta}\right], \label{15}
\end{equation}
and
\begin{equation}
R=16 \pi G \bar{p}.\label{16}
\end{equation}
By evaluating the $g^{\alpha \beta}$ trace of Eq.(\ref{15}) and
combining with Eq.(\ref{16}) we obtain
\begin{equation}
\Box\Phi=\frac{1}{3}(8\pi G(\rho-3p)+16 \pi G \bar{p})\Phi
.\label{18}
\end{equation}
This equation infers the following scalar field potential
\begin{equation}
V(\Phi)=-\frac{1}{6}(8\pi G(\rho-3p)+16 \pi G \bar{p})\Phi^2,
\end{equation}
whose minimum occurs at $\Phi=0$, for which the equations (\ref{15})
reduce to describe a usual $4D$ FRW universe filled with ordinary
matter $\rho$ and $p$. In other words, our conventional $4D$
universe corresponds to the vacuum state of the scalar field $\Phi$.
From Eq.(\ref{18}), one may infer the following replacements for a
nonvanishing $\Phi$
\begin{equation}
\frac{1}{\Phi}\Box \Phi = \frac{1}{3}(8\pi G(\rho-3p)+16 \pi G
\bar{p}),\label{19}
\end{equation}
\begin{equation}
\frac{1}{\Phi}\nabla_{\alpha}\nabla_{\beta}\Phi = \frac{1}{3}(8\pi
G(\rho-3p)+16 \pi G \bar{p})u_{\alpha}u_{\beta}.\label{20}
\end{equation}
Putting the above replacements into Eq.(\ref{15}) leads to
\begin{equation}
G_{\alpha \beta}=8 \pi G [(\rho+\tilde{p})u_{\alpha}
u_{\beta}-\tilde{p}g_{\alpha \beta}], \label{22}
\end{equation}
where
\begin{equation}
\tilde{p}=\frac{1}{3}(\rho+2\bar{p}).\label{23}
\end{equation}
This energy-momentum tensor effectively describes a perfect fluid
with density $\rho$ and pressure $\tilde{p}$. The four dimensional
field equations lead to two independent equations
\begin{equation}
3\frac{\dot{a}^2+k}{a^2}=8 \pi G \rho, \label{24}
\end{equation}
\begin{equation}
\frac{2a\ddot{a}+\dot{a}^2+k}{a^2}=-8 \pi G \tilde{p}. \label{25}
\end{equation}
Differentiating (\ref{24}) and combining with (\ref{25}) we obtain
the conservation equation
\begin{equation}
\frac{d}{dt}(\rho a^3)+\tilde{p}\frac{d}{dt}(a^3)=0. \label{26}
\end{equation}
The equations (\ref{24}) and (\ref{25}) can be used to derive the
acceleration equation
\begin{equation}
\frac{\ddot{a}}{a}=-\frac{4 \pi G}{3}(\rho+3\tilde{p})=-\frac{8 \pi
G}{3}(\rho+\bar{p}). \label{27}
\end{equation}
The acceleration or deceleration of the universe depends on the
negative or positive values of the quantity $(\rho+\bar{p})$. \\From
extra dimensional equation (\ref{16}) ( or $4$-dimensional
Eqs.(\ref{23}), (\ref{24}) and (\ref{25}) ) we obtain
\begin{equation}
-\frac{6(k+\dot{a}^2+\ddot{a}a)}{a^2}=16 \pi G \bar{p}.\label{28}
\end{equation}
Using power law behaviors for the scale factor and dark pressure as
$a(t)=a_0t^{\alpha}$ and $\bar{p}(t)=\bar{p}_0t^{\beta}$ in the
above equation, provided $k=0$ in agreement with observational
constraints, we obtain $\beta=-2$.

Based on homogeneity and isotropy of the 4D universe we may assume
the scalar field to be just a function of time, then the scalar
field equation (\ref{18}) reads as the following form
\begin{equation}
\ddot{\Phi}+3\frac{\dot{a}}{a}\dot{\phi}-\frac{8\pi
G}{3}((\rho-3p)+2 \bar{p})\Phi .\label{29}
\end{equation}
Assuming $\Phi(t)=\Phi_0t^{\gamma}$ and $\rho(t)=\rho_0t^{\delta} \:
(\rho_0>0)$ together with the equations of state for matter pressure
$p=\omega \rho$ and dark pressure $\bar{p}=\Omega \rho$ we continue
to calculate the required parameters for inflation, deceleration and
then acceleration of the universe. In doing so, we rewrite the
acceleration equation (\ref{27}), scalar field equation (\ref{29})
and conservation equation (\ref{26}), respectively, in which the
above assumptions are included as
\begin{equation}
\alpha(\alpha -1)+\frac{8 \pi G}{3}\rho_0(1+\Omega)=0, \label{30}
\end{equation}
\begin{equation}
\gamma(\gamma -1)+3\alpha \gamma-\frac{8\pi
G}{3}\rho_0((1-3\omega)+2 \Omega)=0 ,\label{31}
\end{equation}
\begin{equation}
2\rho_0[(2+\Omega)\alpha-1]=0, \label{32}
\end{equation}
where $\delta=-2$ has been used due to the consistency with the
power law behavior $t^{3\alpha -3}$ in the conservation equation.
The demand for acceleration $\ddot{a}>0$ through Eq.(\ref{27}) with
the assumptions $\rho(t)=\rho_0t^{\delta}$ and $\bar{p}=\Omega
\rho$, requires $\rho_0(1+\Omega)<0$ or $\Omega<-1$ which accounts
for a negative dark pressure. This negative domain of $\Omega$ leads
through the conservation equation (\ref{26}) to $\alpha>1$ which
indicates an accelerating universe as expected.

In both radiation dominant and matter dominant eras
$\omega=\frac{1}{3}$, $\omega=0$, respectively, the scalar field
equation (\ref{31}) leads to the following inequality
\begin{equation}
\gamma(\gamma -1+3\alpha)<0 ,\label{33}
\end{equation}
which through $\alpha>1$ means
\begin{equation}
1-3\alpha<\gamma<0.\label{34}
\end{equation}
It is easily seen in the acceleration equation (\ref{30}) that as
$|\Omega|$ becomes larger and larger, the values of $\alpha$
required for acceleration ($\alpha>1$) become larger and larger, as
well. This result is not surprising, since the more negative dark
pressure we have, the more acceleration is expected. 

On the other hand, using Friedmann equation
we obtain $\alpha=\frac{1}{2+\Omega}$ which together with the
condition $\alpha>1$ requires that  $-2<\Omega<-1$. Now, one may
recognize two options as follows.

The first option is to attribute an intrinsic evolution to the
parameter $\Omega$ along the higher dimension so that it can
produce the $4D$ expansion evolution in agreement with standard
model including early inflation and subsequent deceleration, and
also current acceleration of the universe. Ignoring the
phenomenology of the evolution of the parameter $\Omega$, we may
require
\begin{equation} \label{36}
\left\{ \begin{array}{ll} \Omega \gtrsim -2 \:\:\:\:\:\:\:\:
{for} \:\:\: \mbox{inflation}
\\
{\Omega}>-1 \:\:\:\:\:\:\:\: {for} \:\:\: \mbox{deceleration}
\\
{\Omega}\lesssim-1 \:\:\:\:\:\:\:\: {for} \:\:\:
\mbox{acceleration}.
\end{array}
\right.
\end{equation}
The first case corresponds to highly accelerated universe due to
a large $\alpha>>> 1$. This can be relevant for the inflationary
era if one equate the power law with exponential behavior. The
second case corresponds to a deceleration $\alpha< 1$, and the
third case represents an small acceleration $\alpha \gtrsim 1$.
In this option, there is no specific relation between the
physical phase along extra dimension, namely $\Omega$, and the
ones defined in $4D$ universe by $\omega$. Therefore, an
unexpected acceleration in the {\it ``middle''} of matter
dominated phase $\omega=0$ is justified due to the beginning of a
new phase of $\Omega$.

The second option is to assume a typical relation between the
parameters $\Omega$ and $\omega$ as $\Omega=f(\omega)$ so that
\begin{equation} \label{38}
\left\{ \begin{array}{ll} \Omega \gtrsim -2 \:\:\:\:\:\:\:\:
{for} \:\:\: \omega ={-1}
\\
{\Omega}>-1 \:\:\:\:\:\:\:\: {for} \:\:\: \omega=\frac{1}{3}
\\
{\Omega}\lesssim-1 \:\:\:\:\:\:\:\: {for} \:\:\: \omega=0.
\end{array}
\right.
\end{equation}
The physics of $\omega$ is well known in the standard cosmology (see
bellow) but that of the parameter $\Omega$ clearly needs more
careful investigation based on effective representation of higher
dimensional theories, for instance string or Brane theory. In fact,
at early universe when it is of Plank size, it is plausible that all
coordinates including 3-space and higher dimension would be
symmetric with the same size. However, during the GUT era when some
spontaneous symmetry breakings could have happened to trigger the
inflation, one may assume that such symmetry breakings could lead to
asymmetric compact 3-space and non-compact higher dimension. This
could also result in demarcation between three spatial pressures on
3-space and one higher dimensional pressure. Therefore, it is
possible to consider a relation $\Omega=f(\omega)$ which is based on
the physics of early universe (phase transitions) along with initial
conditions to justify this demarcation.

The case $\omega = -1$  corresponds to the early universe and shows
a very high acceleration due to $\alpha>>> 1$. The case
$\omega=\frac{1}{3}$ corresponds to the radiation dominant era and
shows a deceleration $\alpha< 1$. Finally, the case $\omega=0$
corresponds to the matter dominant era and shows an small
acceleration $\alpha \gtrsim 1$ at the {\it ``beginning''} of this
era.

\newpage
\section*{CONCLUSION}

A $(4+1)$-dimensional universe consisting of a $(4+1)$ dimensional
metric of Robertson-Walker type subject to a $(4+1)$ dimensional
energy-momentum tensor in the framework of noncompact Kaluza-Klein
theory is studied. In the matter part, there is energy density
$\rho$ together with pressure $p$ subject to $4D$ part of the
$(4+1)$ dimensional energy-momentum tensor, and a dark pressure
$\bar{p}$ corresponding to the extra-dimensional part endowed by a
scalar field. The reduced $4D$ and extra-dimensional components of
$5D$ Einstein equations together with different equations of state
for pressure $p$ and dark pressure $\bar{p}$ may lead to a $4D$
universe which represents inflation for early universe, deceleration
for radiation dominant and acceleration for mater dominant eras.
This is done by assuming a typical relation (\ref{35}) between the
two parameters in equations of state. This relation is not unique
and one may propose other suitable relations, satisfying the
requirements for early inflation, then deceleration and finally
recent acceleration of the universe in a more realistic way. The
important point of the present model is that there is no longer
``coincidence problem''. This is because, in the present model there
is no element of ``dark energy'' at all and we have just one energy
density $\rho$ associated with ordinary matter. So, there is no
notion of coincidental domination of dark energy over matter
densities to trigger the acceleration at the present status of the
universe. In fact, a dark pressure with negative values have existed
along the $5^{th}$ dimension for the whole history of the $4D$
universe including inflationary, radiation, and matter dominant
eras. These stages of the 4$D$ universe have occurred because of
negative, positive and zero values of the four dimensional pressure,
respectively, which leads through relation (\ref{35}) to a
competition between energy density $\rho$ and dark pressure
$\bar{p}$ in the acceleration equation (\ref{27}). For the same
reason that there is no element of dark energy in this model, the
apparent {\it phantom like} equation of state for dark pressure
$\Omega<-1$ is free of serious problems like {\it unbounded from
below dark energy} or {\it vacuum instability} \cite{Cline}.

The above results are independent of the signature $\epsilon$ by
which the higher dimension takes part in the 5D metric. Moreover,
the role played by the scalar field along the $5^{th}$ coordinate in
the $5D$ metric is very impressed by the role of scale factor over
the $4D$ universe. At early universe during the inflationary era the
scalar field is highly suppressed and the $5^{th}$ coordinate is
basically ignored in $5D$ line element. At radiation dominant era
the scalar field is much less suppressed and the $5^{th}$ coordinate
becomes considerable in $5D$ line element. Finally, at matter
dominant era the scalar field and its possible fluctuations starts
to be more suppressed and the observable effect of $5^{th}$
coordinate becomes vanishing in $5D$ line element at $t\simeq
10^{17}{Sec}$, leaving a $4D$ universe in agreement with
observations.

\end{document}